# AN OVERVIEW OF SOME INFLUENTIAL FACTORS IN LASER INDUCED BREAKDOWN SPECTROSCOPY

**V. N. Rai**

**Raja Ramanna Centre for Advanced Technology**

**Indore- 452013 (India)**

**E-mail: vnrai@rrcat.gov.in**

**Phone- 0731-2488142**

**ABSTRACT**

This paper presents the process of plasma formation during laser induced breakdown spectroscopy (LIBS) along with analysis of emission from double pulse LIBS to better understand the processes and the factors involved in enhancement of plasma emission. In this analysis plasma emission has been considered directly proportional to the square of plasma density, its volume, and the fraction of absorption of second laser pulse in the plasma plume produced by the first laser through inverse Bremsstrahlung absorption process. The electron–ion collision frequency, which is dependent on the density and temperature of the plasma, has been found playing important role in the enhancement of emission as well as in the saturation of emission during LIBS. The effect of material ablation, delay between lasers, plasma confinement and shielding effect has also been discussed.

## 1. INTRODUCTION

Laser-induced breakdown spectroscopy (LIBS) is a spectroscopic technique utilizing high power laser pulses [1-5]. When a pulsed high power laser beam is focused on the material (Solid, Liquid, Gas and aerosols), a short lived (micro second) high temperature (few eV) and density ($10^{16}$- $10^{19}$ $cm^{-3}$) plasma is produced and expands perpendicular to the target surface. Such laser produced plasmas, which contains a diversity of atomic and ionic species as well as free electron emit radiation over an extensive spectral range stretching from infrared to the X-ray region. Although different materials have different breakdown thresholds, the plasma is typically generated when the laser irradiance exceeds ~ $10^8$ $MW/cm^2$. The radiation emitted from excited/ionized species in the laser produced plasma is spectrally resolved, and the individual neutral atoms and/or atomic ions in the plume are identified and quantified by their unique wavelengths and the line intensities respectively. The fundamental concept of LIBS is similar to that of conventional plasma-based methods of optical emission spectroscopy (OES). In practice this technique is calibrated through establishing a relationship between the observed spectral line intensities of a particular species and the corresponding concentration in the target sample material. The graph that corresponds to this relationship is called the calibration curve and is used to determine the elemental contents of unknown sample by recording its LIBS intensity. Since its starting LIBS technique has been employed for the direct qualitative as well as quantitative elemental characterization of almost every type of material using optical emission spectroscopy. There are many reviews and books dedicated to basic principle and instrumentation of LIBS as well as to a wide range of applications [1-9].

In the context of spectrochemical analysis of elements, LIBS has many advantages over other conventional spectroscopic techniques such as laser ablation inductively coupled plasma (LA-ICP), X-ray fluorescence (XRF), atomic absorption spectroscopy (AAS) and spark-discharge optical emission spectroscopy (SD-OES). These advantages generally include the fact that sample preparation is either not necessary or minimal. Sample excitation/ionization in the LIBS is carried out by optical energy. It is an almost non-destructive technique that provides direct characterization of the solid sample. The most important point about LIBS is that it has powerful capability in carrying out remote on line and in-situ analysis of the samples particularly situated in the hostile and harsh environments. However it is lacking on the sensitivity front. Various techniques have been used to enhance the sensitivity of the LIBS. Some of them include expansion of plasma in the magnetic field, dual and multiple pulse excitation of plasma, spatial confinement of LIBS plasma as well as combination of LIBS with laser induced fluorescence technique [8]. Dual pulse excitation technique has been widely studied experimentally as well as theoretically to enhance the sensitivity of the LIBS as well as to better understand the important factors effecting enhancement in plasma emission [10-13].

This paper presents an overview of the plasma formation as a result of laser matter interaction along with an enhancement in plasma emission under double pulse excitation as well as the most important factors affecting it. Mainly effects of plasma parameters such as the plasma density and temperature as well as the laser pulse duration affecting the dynamics of plasma and its emission characteristics are also discussed. It will be helpful in better understanding the physical processes taking place in the plasma during LIBS as well as in increasing its sensitivity.

## 2.    PROCESS OF PLASMA FORMATION IN LIBS

LIBS is a very simple spectroscopic technique, which uses a high power laser beam focused on the sample to produce plasma. The emitted light is collected through an optical fiber and diverted into a spectrometer. The physics and chemistry of plasma initiation, formation along with its life time and decay are very complicated [14]. Much progress has been made in the study of plasma generation processes, which includes local thermodynamic equilibrium (LTE) models, hydrodynamics and kinetic models, non-uniform plasmas and plasma generated in the vacuum [15-17].

There are two main processes that play important role in ionization of atomic and molecular species in laser-induced breakdown. The first one is the direct ionization of the sample by multiphoton ionization (MPI), whereas the second one is the inverse bremsstrahlung (IB) absorption processes [16, 18-20]. In MPI atoms or molecules undergo simultaneous absorption of sufficient number of photons to cause ionization (or the rejection of electrons from the valence to conduction band in the case of metals) as

$$M + m\ (h\nu) \longrightarrow M^+ + e^-$$ -------------- (1)

where $\varepsilon_1$ is the ionization potential and m is the number of photons, which must be greater than the integral number of $(\varepsilon_1/h\nu +1)$. MPI is significant at wavelengths shorter than $\sim 1$ μm and at high laser power that is greater than $10^{10}$ W/cm$^2$. At wavelength substantially longer than $\sim 1$ μm, it is statistically unlikely for an atom or molecule to absorb enough photons to increase the energy of the neutrals above its ionization potential. However this process is important at low pressure, when collision frequencies between species are low due to the low particle density of the medium.

IB absorption processes involve the absorption of a photon by one or more seed electrons present in the focal volume at the beginning of the laser pulse. The initial free electrons may be generated through the presence of cosmic rays. The first few photons in the laser pulse can also produce seed electrons from dust, $O_2$ as well as organic vapours. It may also come from atoms and molecules present in the atmosphere via MPI. Seed electrons are not necessary for pure MPI processes.

In IB the absorption of photon increases the electron energy to a higher state in the continuum. This process is possible in the vicinity of heavy particles, such as an atom, ion or molecule so that momentum is conserved. In normal bremsstrahlung processes, high energy electrons emit radiation as they slow down upon interacting with heavy particles. In the case of collision electrons lose its energy by rotational and vibrational excitation of neutral molecules, excitation of electronic states of the atoms and molecules as well as by attachment of electrons. In the case of IB, electrons acquire energy from the absorbance of photons in the presence of collision with atoms, ions and molecules. Once the energy of the free electron is greater than the ionization potential of the neutral species, it can ionize atoms or molecule (M) by colliding with it. This produces two lower energy free electrons, which can gain more energy from the electric field, causing ionization of other neutrals and two more electrons as

$$e + M \longrightarrow 2e^- + M^+$$ ------------ (2)

An increase in the population of electrons and ions in the focal volume, the probability of electron-photon neutral collisions also increases, resulting in electron multiplication and cascade growth. Repeated increase in the number of free electrons result in a cascade ionization. During IB processes, all the species ablated from the target sample can be ionized to form plasma and the entire laser pulse energy can be coupled into the plasma. This results in an increase in density of the plasma in such a way that it becomes optically opaque. This is because laser can not propagate in the plasma having density more than the critical density (where plasma frequency becomes equal to the laser frequency). In such a situation sample material becomes shielded from remainder of the laser pulse.

The plasma producing processes such as MPI and IB absorption mainly depend on the wavelength of the laser radiation, laser intensity and density of the medium in which the laser breakdown occurs. IB dominated breakdown is important at high pressures, when collisional effects are strong and at wavelength longer than 1 μm. At shorter irradiation wavelength ($\leq 1\mu m$) or at low densities of molecules, the possibility of electrons colliding with neutral species is small. Therefore, MPI dominates usually at these shorter wavelengths and in low density media. Finally cascade ionization continues throughout the presence of the laser pulse resulting in the ionization and dielectric breakdown of gases as well as vaporized particles creating plasma.

## 3.    EMISSION FROM DOUBLE PULSE LIBS

Normally double pulse LIBS [11] uses two laser pulses separated by microsecond time delay. The first laser pulse creates plasma expanding normal to the target surface, whereas the second laser pulse gets absorbed in it followed by further ablation of target material by the residual second laser pulse. The absorption of second laser in the plasma created by the first laser pulse can be dominated by inverse Bremsstrahlung absorption, which occurs due to electron – ion collisions in the plasma as discussed in the previous section. Once plasma electrons are subjected to momentum changing collisions as they oscillate back and forth in the laser electric field, the laser light wave undergoes an effective damping. The spatial damping rate of wave energy, $k_{ib}$ can be given by [16]

$$k_{ib} \propto \frac{Z n_e^2}{T_e^{3/2} \left(1 - \dfrac{n_e}{n_c}\right)^{1/2}} \tag{3}$$

Where $n_e$ and $T_e$ is the density and temperature of the plasma and Z is the atomic number of the target material. This shows that inverse Bremsstrahlung is strongest for the low temperature, high density and high Z target materials. According to eq. (3) a uniform plasma of length L and density $n_e < n_c$ will produce a one way absorption fraction given by

$$\alpha_{abs} = 1 - \exp\left(-k_{ib}L\right) \qquad (4)$$

For weak absorption $k_{ib}L \ll 1$ and $\alpha_{abs} \cong k_{ib}L$. For strong absorption, $k_{ib}L \gg 1$ and $\alpha_{abs} \rightarrow 1$. This indicates that absorption fraction is linear in $k_{ib}L$ for small absorption and then saturates, when $k_{ib}L$ is large. Even the long plasma length enhances the absorption. The result of inverse Bremsstrahlung absorption in inhomogeneous plasma is more complicated. A simplified version of absorption fraction for linear density profile is given by [16]

$$\alpha_{abs} = 1 - \exp\left\{-\frac{32}{15} \frac{L \nu_{ei}(L)}{c}\right\} \qquad (5)$$

where $\nu_{ei}$ (L) is the electron-ion collision frequency at critical density and c is the speed of light. This indicates that inverse Bremsstrahlung absorption is not only strongest for the low temperature, high densities, high Z targets but for the long density scale length plasma also. It is sensitive to details of the density profile near the critical surface and can be diminished considerably if the profile falls steeply at the critical surface. For simplification purpose, one can consider plasma expansion length $L \approx C_sT$, where $C_s$ is the ion acoustic (plasma expansion) speed and T is the plasma expansion time. An expression for electron–ion collision frequency $\nu_{ei}$ can be given as [21]

$$\nu_{ei} = 3.62 \times 10^{-6} n_i \, T_e^{-3/2} \ln\Lambda \ \ \text{s}^{-1} \qquad (6)$$

where $n_i$ ($m^{-3}$) is the ion density in the plasma, $T_e$ (K) is the plasma temperature and $\ln\Lambda \sim 10$. The electron–ion collision frequency $\nu_{ei}$ is considered to be the constant over the whole plasma length.

  The plasma expansion in planar geometry can be considered in the form of cylinder. Therefore volume of the radiating plasma can be approximated by $V = \pi R^2 L$ for a plasma column of diameter 2R and length L. The mass ablated by the first and the second laser pulse can be given as $\dot{m}_1 \pi R^2 \tau_L$ and $\dot{m}_2 \pi R^2 \tau_L$ respectively, where $\dot{m}_1$ and $\dot{m}_2$ are the mass ablation rate after the first and the second laser pulse and $\tau_L$ is the time duration of both the lasers. All the parameters of both the lasers are considered to be the same. It is well known that the plasma emission takes place through three processes as Bremsstrahlung emission, free-bound transition and bound-bound transition. In each case, the plasma emission is strongly dependent on the product of electron and ion density ($\propto$ $n_e n_i$), the temperature of plasma as well as on the volume of plasma. Considering that the plasma emission is proportional to the square of the plasma density ($n_e = n_i$) and the plasma volume, one can write an expression for the emission intensity ($I_1$) after first laser pulse as

$$I_1 \quad \propto \quad \left( \frac{\dot{m}_1 \pi R^2 \tau_L}{\pi R^2 C_{s1} T_1} \right)^2 \left( \pi R^2 C_{s1} T_1 \right) \tag{7}$$

Where first factor is the plasma density and the second factor is volume of the plasma. $C_{s1}$ is the ion acoustic speed of the plasma due to first laser pulse and $T_1$ is the time of plasma expansion that is gate delay from the first laser, when the emission is recorded in single pulse LIBS.

Emission after the second laser pulse ($I_2$) can be written as

$$I_2 \quad \propto \quad \left( \frac{\dot{m}_2 \pi R^2 \tau_L}{\pi R^2 C_{s1} \Delta t + \pi R^2 C_{s2} T_2} \right)^2 \left( \pi R^2 C_{s1} \Delta t + \pi R^2 C_{s2} T_2 \right) \left[ 1 - \exp \left\{ -\frac{32}{15} \frac{C_{s1} \Delta t \, \nu_{ei}}{c} \right\} \right] \quad \text{---} \quad (8)$$

Here the first factor describes the plasma density after the time ($\Delta t + T_2$), the second factor represents the plasma volume after the time ($\Delta t + T_2$) and the third factor equals the fraction of the second laser absorbed in the plume of the plasma created by the first laser pulse. Here $\Delta t$ is the time delay between the lasers, $C_{s2}$ describes the ion acoustic speed after the second laser pulse and $T_2$ (gate delay) is the time when recording of the plasma emission takes place after firing the second laser. To deduce a simplified expression for enhancement in double pulse emission and canceling all the common factors affecting the single and double pulse LIBS emission, the enhancement (E) is obtained by normalizing the sum of plasma emission under single and dual pulse LIBS by emission from single pulse LIBS [22].

$$E = \frac{I_1 + I_2}{I_1} \quad = 1 + \frac{I_2}{I_1}$$

$$E = 1 + \left(\frac{\dot{m}_2}{\dot{m}_1}\right)^2 \frac{\left[1 - \exp\left\{-\frac{32}{15}\frac{C_{s1}\Delta t\, \nu_{ei}}{c}\right\}\right]}{\left(\frac{\Delta t}{T_1} + \frac{C_{s2}}{C_{s1}}\frac{T_2}{T_1}\right)} \qquad ---- \qquad (9)$$

where one can take for the value of $T_1$ any value between firing of second laser and the time of peak emission in double pulse LIBS without any significant change in the result. Eq. (9) has been used to study the effect of different parameters on the enhancement in plasma emission under dual pulse excitation, which explained successfully the various observations of dual pulse LIBS as reported earlier [22]. According to eq. (9) zero delay between two lasers ($\Delta t = 0$) provides an enhancement of unity ($E = 1$). It is just like the case of a single pulse LIBS emission with added laser intensity.

The above eq.-9 obtained for dual pulse excitation of the plasma can be simplified for single pulse case under the consideration that the plasma is formed and starts expanding when the initial portion of laser pulse interacts with the matter, whereas rest of the pulse will be absorbed in the preformed plasma. So a single laser pulse can also be considered operating as dual pulse with certain assumptions. Mainly one can consider that after the plasma formation by pedestal of the laser, it expands for the period $\tau_L$ (duration of laser pulse) during which laser gets absorbed in the expanding plasma. This indicates that $\Delta t$ delay between two lasers in eq.-9 can be replaced by $\tau_L$ for the single pulse case. The eq.-9 can be simplified in two ways as described below to better understand the dependence of important plasma parameters on the emission characteristics of the plasma as well as on its temporal behavior.

**Case - 1**

In this case assumption is made that $\Delta t = \tau_L$ (laser time duration), $T_1 = \tau_L/2$ (time of peak laser intensity). $T_2$ is the time at which measurement of the plasma emission is taking place. Other consideration is that $C_{s2} = 2\, C_{s1}$. In this situation eq.-9 can be written as

$$E = 1 + \frac{\left(\dfrac{\dot{m}_2}{\dot{m}_1}\right)^2 \left[1 - \exp\left\{-\dfrac{32}{15}\dfrac{C_{s1}\tau_L \nu_{ei}}{c}\right\}\right]}{\left(2 + 4\dfrac{T_2}{\tau_L}\right)} \qquad \text{-------} \qquad (10)$$

This equation can provide information about the emission from the plasma at different time duration that means the variation of emission with the gate delay ($T_2$).

**Case – 2**

In this case we consider $\Delta t = \tau_L$ and $T_1 = T_2 = T$ is the gate delay or the time of measurement of the emission after the laser peak. Ratio of the laser pulse time duration $\tau_L$ and gate delay T (time of measurement) will be negligibly small for nanosecond time duration laser ($\tau_L / T \approx 0$). After these assumption eq.-9 can be written as

$$E = 1 + \left(\frac{\dot{m}_2}{\dot{m}_1}\right)^2 \left(\frac{C_{s1}}{C_{s2}}\right)\left[1 - \exp\left\{-\frac{32}{15}\frac{C_{s1}\tau_L \nu_{ei}}{c}\right\}\right] \qquad \text{----------} \qquad (11)$$

This indicates that the plasma emission under single pulse laser plasma interaction is mainly dependent on the square of the ratio of the mass ablation rate after full pulse and initial mass ablation, the ratio of plasma sound velocity, the plasma collision frequency and pulse duration of the laser.

## 4. RESULTS AND DISCUSSIONS

In order to achieve a better understanding of the physical processes arising in double pulse LIBS, calculations are made for the enhancement in emission for different plasma parameters such as mass ablation rate, delay between lasers, gate delay, plasma density and temperature along with electron ion collision frequency.

### 4.1 Effect of Delay Between Lasers on LIBS Emission

The delay between two lasers is an important parameter, which decides an optimum enhancement in plasma emission. The enhancement in plasma emission has been estimated for different values of inter-pulse delay between lasers ($\Delta t$) in the range from 0 to 30 µs using Eq. (9). For simplification an assumption has been made that mass ablation rate and plasma expansion speed in single and double pulse LIBS remains nearly same such that $\dot{m}_1 = \dot{m}_2$ and $C_{s1} = C_{s2}$. Here $t_1$ and $t_2$ are taken as 4 and 2 µs respectively, which implies that in the case of single pulse LIBS, emission is recorded after 4 µs (for normalization), whereas in double pulse, LIBS emission is measured ~ 2 µs after firing the second laser. The variation in emission enhancement was calculated for three values of electron–ion collision frequency $\nu_{ei} = 3.62 \times 10^9$, $1 \times 10^{10}$ and $3.62 \times 10^{10}$ s$^{-1}$. Fig.-1 shows the variation in enhancement E with change in delay between the two lasers (Δt) for different electron-ion collision frequency. The enhancement remains unity for Δt = 0 whereas it increases sharply and attains a peak followed by a slow decrease with an increasing Δt. The peak of the emission shifts to lower delay as the electron–ion collision frequency increases from $3.62 \times 10^9$ to $3.62 \times 10^{10}$. Enhancement is nearly ~ 2 times in the case of $\nu_{ei} \sim 3.62 \times 10^{10}$ s$^{-1}$ for Δt ~2 µs. The enhancement intensity is found to be less and emission profile is broadened for lower values of electron–ion collision frequencies. Higher electron–ion collision frequency corresponds to high plasma density and low plasma temperature, which is possible with a moderate intensity laser creating low temperature plasma, whereas interaction with second laser takes place before its proper expansion. In case second laser pulse interacts with the low density plasma already expanded for > 2 µs time delay, it gets absorbed in longer path length resulting in a wide emission profile peaking at comparatively higher inter-pulse delay with decreased emission intensity. Similar behavior is possible, when a high intensity laser is used as first laser for producing pre-plasma. In such case plasma expands faster as a result of high temperature providing a broad density profile. The interaction of the second laser with such plasma will provide less enhancement in emission with a broadened profile peaking at comparatively higher inter pulse delay. The maximum emission (peak) occurs for inter pulse delay (Δt) of 1, 3 and 5 µs for electron-ion collision frequencies of ~$3.62 \times 10^{10}$, $1 \times 10^{10}$ and $3.62 \times 10^9$ s$^{-1}$, with the corresponding enhancements of 1.96, 1.52 and 1.27 respectively (Fig.-1).

These theoretically obtained results are found in agreement with experimental observations reported by Rai et al. [11], which shows the emission from magnesium (Mg) ion (279.55 nm) recorded in double pulse excitation. It is found that emission profile peak observed at inter-pulse delay of 2-3 µs corresponds to an enhancement of ~ 6 times, where the first laser energy was kept at ~100 mJ (moderate intensity). However as the energy of first laser increases from 100 to 180 mJ, the emission profile is broadened with a peak at an inter-pulse delay more than 5 µs.  The theoretically computed emission profile peaks at ~ 3 µs inter-pulse delay for electron-ion collision frequency of ~$1 \times 10^{10}$ s$^{-1}$ with an enhancement of E ~ 1.52, whereas this peak shifts to ~ 5 µs inter-pulse delay for electron-ion collision frequency of~ $3.62 \times 10^9$ s$^{-1}$ with an enhancement of E ~1.27.  The large difference in the amplitude of enhanced emission in theoretical and experimental results may be dependent on the other factors, which will be discussed in the next section.

It is relevant to mention here that inter-pulse delay between lasers for peak emission enhancement varies in different experiments depending on the experimental conditions as reported earlier. Forsman et al. [24] reported that several laser pulses separated by several tens of nanoseconds substantially enhance plasma emission from 3-10 times, which was correlated by enhanced material ablation. Uebbing et al. [25] obtained enhancement in emission intensity for inter-pulse delay time between $5 - 50$ µs and explained it as reheating of the plasma by the second laser-pulse. Kuwako et al. [26] also reported an optimum inter pulse delay time of ~ 8 µs for detection of sodium in water. Burakov et al. [27]  found enhancement in emission bands of $C_2$ and $C_3$ for inter pulse delay between $0 - 5$ µs. An increase in intensity of $C_2$ was observed for 2 µs inter-pulse delay time, whereas increase in enhancement of $C_3$ emission was noted for inter-pulse delay time up to 5 µs. Enhancement effect for $C_2$ emission was lasted up to ~30 µs. Around ~ 33 fold signal enhancement was observed by Stratis et al. [28-29] with the use of combination of air spark and a laser pulse separated by 2.5 µs delay. Similarly Mao et al.[30]  studied the effect of inter-pulse delay ranging from $1 - 10$ ns between nanosecond time duration laser pulses on the plasma temperature, emission intensity, particle number density and crater volume. Rai et al. [31] have reported that optimum absorption of second laser in pre-plasma is possible only when the plasma expansion length $C_s. \Delta t >> \lambda$ (wavelength of the laser). These observations in addition to various other results indicate that optimum inter-pulse delay for maximum enhancement in emission depends on various parameters such as experimental configuration, parameters of the laser used for producing plasma as well as on the properties of the target materials being ablated.

## 4.2    Role of Material Ablation on LIBS Emission

An enhancement in emission has been calculated theoretically for two values of ratio of mass ablation rate in double and single pulse LIBS ($\dot{m}_2 / \dot{m}_1$) such as 1 and 3.5 for $\nu_{ei} \sim 1 \times 10^{10}$ s$^{-1}$. The variation in enhancement in emission with delay between lasers is shown in Fig.-2 for different ($\dot{m}_2 / \dot{m}_1$). Maximum enhancement in emission is found to be ~7 and 1.52 times for $\dot{m}_2 / \dot{m}_1 = 3.5$ and 1 respectively (Fig.-2). Nearly ~7 times enhancement in emission is in close agreement with the experimental observation reported earlier, which indicates that rate of material ablation increases $\geq$ 3.5 times in the case of double pulse in comparison to single pulse LIBS [11]. Benedetti et al. [32] have also reported a correlation between an increase in ablated mass and enhancement in the LIBS signal under double pulse excitation from solid target. They found that crater produced in target material in double pulse configuration is much deeper than that produced by a single pulse laser. The volume of the crater in double pulse LIBS was found larger than that obtained in single pulse LIBS by a factor of 4 - 6.

Experimental findings of various other groups are also in qualitative agreement with the theoretical results that material ablation increases manifold after second laser pulse. The possible explanation for such an enhancement in material ablation has also been discussed. Koudoumas et al. [33] has found that enhancement is related to the creation of a melted zone on the target by the first laser with modified optical coupling properties, where second laser pulse interacts with a liquid surface having different optical properties leading to total vaporization of the melted front leaving behind a smooth structure. Spyridaki et al. [34] found that the optimal separation between pulses corresponds to the formation of a liquid layer of thickness equal to the optical penetration depth leading to the formation of smooth structure on the surface. Stoian et al. [34] indicated that the sequential energy delivery induces softening in target material during the initial laser pulse changing the energy coupling for the subsequent steps. The most probable reason for emission enhancement in double pulse LIBS was found related to an improved laser ablation by the second laser pulse due to the rarefied gas density inside the region encompassed by the shock wave produced by the first laser pulse [36]. Another suggestion about the increase in the emission during double pulse LIBS is due to an additional increase in the concentration of emitting species due to the vaporization of large particles (clusters) formed after the first laser pulse. Gonzalez et al. [37] suggested that increase in emission intensity (Cu, Zn) originates from the vaporization of particles (large clusters) by the second pulse. Burakov et al.[27] indicated that emission enhancement in $C_2$ and $C_3$ may be related to the production of additional $C_2$ and $C_3$ molecules through the decomposition of carbon clusters evaporated in particulates form after the first laser pulse. Observation of cluster formation in liquid samples has also been indicated in double pulse LIBS experiment reported earlier by Rai et al. [11].

## 4.3 Temporal Evolution of LIBS Emission

Gate delay is an important parameter, which provides information about the dynamics of the plasma along with decay in the plasma emission after its formation as it expands away from the target surface. The effect of gate delay on the enhancement of plasma emission under double pulse excitation is studied using eq. (9). For this purpose the delay between lasers is fixed as $\Delta t = 3$ μs and the ratio of mass ablation rate $\dot{m}_2 / \dot{m}_1 =$ 3.5, which is best suited with the experimental result [11]. An assumption is made here that plasma temperature remains nearly same in single and double pulse LIBS configuration resulting nearly same plasma expansion speed, that is, $C_{s1} = C_{s2}$. The calculation is made for two values of electron-ion collision frequency $\nu_{ei} = 1 \times 10^{10}$ and $3.62 \times 10^{11}$ s$^{-1}$ and $t_1$ is taken as 4 μs the gate delay after first laser pulse (used for normalization purpose). Fig.-3 shows the variation in emission enhancement with an increase in gate delay. Maximum enhancements of ~ 17 and 12 times are found for $\nu_{ei}$ of $3.62 \times 10^{11}$ and $1 \times 10^{10}$ s$^{-1}$ respectively, when the gate delay after the second pulse $t_2$ is taken as zero. However experimental data for $t_2 = 0$ (just after firing the second laser) is not possible, because of high intensity of Bremsstrahlung emission, where detectors show saturation. Detection of emission becomes possible only at $t_2 = 2$ μs after the significant decay of Bremsstrahlung (background) emission [11]. The enhancement in emission initially decreases very fast and then slowly with an increase in the gate delay after second laser pulse. This is possible due to fast expansion of plasma after the second laser pulse as a result of the plasma cooling along with the loss of emitting species out of observation area due to diffusion process in the plasma.

The above discussed theoretical results are found in agreement with the experimental findings of double pulse LIBS emission after 2 μs gate delay (from second laser) [11]. The maximum enhancement in intensity of double pulse LIBS at 2 μs gate delay with respect to single pulse LIBS emission at 4 μs gate delay was recorded ≥ 12 times. At this point the line emission intensity is at maximum and starts decaying with an increase in the gate delay. These experimental observations are in qualitative agreement with the theoretical results discussed earlier. The agreement of experimental observations with theoretical calculation for maximum enhancement in emission of ≥ 12 times, further supports that our assumption of $\dot{m}_2 / \dot{m}_1 = 3.5$ is nearly correct. This indicates that enhancement in emission under double pulse LIBS [11] is mainly due to an increase in mass ablation rate by ≥ 3.5 times after second laser pulse for given experimental conditions [22].

### 4.4    Effect of Plasma Volume and Plasma Confinement on Emission

It is important to discuss the effect of temperature and the volume of plasma on the enhancement in emission under double pulse LIBS experiment. Eq. (9) has been used to calculate enhancement (E) for electron-ion collision frequency $\nu_{ei} \sim 3.62 \times 10^{10}$ s$^{-1}$ and $\dot{m}_1 = \dot{m}_2$, which corresponds to the same value of mass ablation rate induced by the first and the second laser pulse. An increase in plasma temperature / plasma volume is related with the plasma expansion speed (ion acoustic speed). Fig.-4 shows the variation in enhancement in emission (E) for $C_{s2}/C_{s1} = 1$ and 1.5 with change in delay between the lasers. The higher value of $C_{s2}$ in comparison to $C_{s1}$ indicates increase in the plasma volume, whereas decrease signifies plasma confinement. Enhancement in emission is more for $C_{s2}/C_{s1} = 1$ in comparison to $C_{s2}/C_{s1} = 1.5$, which is expected, because the increased plasma volume has same number of emitting species. This is just the case of decrease in the concentration of emitting species. An increase in ratio of mass ablation rate to $\dot{m}_2 / \dot{m}_1 = 3.5$ increases the enhancement in emission even if $C_{s2}/C_{s1} \sim 1.5$. This simply indicates that an increase in plasma temperature and related plasma expansion speed/plasma volume adversely affects the enhancement in emission. However an increase in plasma volume can contribute to enhancement in emission only when the density of emitting species increases through enhanced ablation by the second laser. It has been reported [23] that the emission can be increased by confining the plasma that is by decreasing the value of $C_{s2}$ / $C_{s1}$ to less than one as shown in Fig.-5. Various techniques of plasma confinement such as magnetic or spatial confinement have been used and reported to enhance the plasma emission by manifold [8, 38]. This suggests that combination of emission enhancement process can increase the emission intensity manifold and consequently sensitivity of LIBS technique [23, 38].

## 4.5    Saturation Effect in Plasma Emission

The study of variations in plasma emission with electron density and temperature for laser of different time duration can provide better understanding of the  saturation in plasma emission. For this purpose calculation is performed using eq.-11 with a consideration that $\dfrac{\dot{m}_2}{\dot{m}_1} = 4,$     $\dfrac{C_{s2}}{C_{s1}} = 2$ and $T_e = 1 \text{eV}$ and the results are presented in Fig.-6. This shows variation in plasma emission (E) with change in plasma density for different time duration of the laser $\tau_L$. It is shown that the plasma emission starts increasing at $n_e \sim 10^{16} \text{ cm}^{-3}$ for a laser of 1 ns time duration. This increase became fast with an increase in plasma density, which saturates at $\sim 10^{19} \text{ cm}^{-3}$. Similar variations are  noted for 10 and 20 ns time duration laser, but with minor change in the threshold for initiation and saturation of emission, which occur at lower density for large time duration lasers. Higher time duration laser produce comparatively more plasma emission as a result of more material ablation [39-40]. Here the electron-ion collision frequency and the laser plasma interaction time play an important role. In this case plasma collision frequency increases with an increase in the plasma density, which helps in increasing absorption of the laser in the plasma through inverse Bremsstrahlung and consequently the plasma emission, whereas large time duration of laser provides longer time for plasma to expand and absorb the laser efficiently.

The effect of plasma temperature on its emission has also been obtained using eq.-11. Here plasma density is considered constant at ~$10^{17}$ cm$^{-3}$ keeping all the other parameters same as in Fig.-6. In this case emission from plasma is calculated by varying the plasma temperature for different time duration laser. Fig.-7 shows that plasma emission is very high for low temperature plasma, which decreases first very fast and then slowly with an increase in plasma temperature. The reason for decrease in plasma emission with an increase in plasma temperature is due to decrease in laser absorption in plasma as a result of decrease in electron-ion collisions frequency as $v_{ei} \propto (T_e)^{-3/2}$. Here again plasma emission is high for longer time duration laser as seen in earlier case. In this case saturation is seen at lower plasma temperature near ~ 1 eV. Results of Fig.- 6 & 7 indicates that saturation in the plasma emission is possible only when the plasma density is high and plasma temperature is low. In other words one can say that instead of plasma density and plasma temperature, electron-ion collision frequency play an important role in deciding the saturation in plasma emission, which occurs at $v_{ei} > 10^{13}$ s$^{-1}$ as has been discussed earlier [23].

### 4.6  Saturation Due to Shielding Effect in the Plasma

The saturation in LIBS emission can also be understood in a different way by considering variation in the laser absorption mechanism with an increase in laser intensity. For lower laser intensity a laser-supported detonation wave (LSDW) is observed in the plasma where the laser absorption region propagates as a shock wave [8, 41]. In this regime if the electron density remains below the critical density (where plasma frequency is equal to laser frequency) then absorption of laser intensity will increase in the plasma resulting in more ablation of target material and consequently the plasma emission. This is the reason why plasma emission increases linearly with an increase in laser intensity. When the plasma density reaches near critical density level, the plasma becomes opaque to the laser beam, which shields the target. In this case coupling of laser intensity to plasma decreases, which leads to a saturation in plasma emission. This phenomenon can be observed in both types of targets (solid and liquid) when the laser intensity is in the medium range. This saturation takes place at comparatively low laser intensity for solid in comparison to  liquid sample due to production of comparatively higher plasma density in the case of solid targets. At higher laser intensity absorption mechanism changes to laser-supported radiation wave, in which plasma temperature becomes very high resulting in a transparent plasma for the laser beam, which starts emitting Bremsstrahlung (Background) emission due to its high temperature (Aguilera et al., 1998). The plasma shielding effect is then reduced and the ablation of the target material enhanced. In this regime also plasma emission (x-rays and visible) intensity increases linearly. A saturation in the x-ray emission has also been noticed in the form of change in the slope with an increase in the laser intensity. However this saturation may have contribution from the loss of plasma energy in the form of generation of different kinds of plasma instability and high energy particles in the plasma along with emission of high energy x-rays [42-43]. Normally high intensity laser experiment involves measurement of x-ray emission in vacuum condition. However high laser intensity experiment in air also shows saturation, where shielding of laser occurs by air plasma [41, 44]. In this situation, presence of solid target surface near laser focus enhances breakdown in the surrounding gas, which is why air plasma is also formed when the laser is focused below or on the sample surface [45].

## 5.     CONCLUSION

In summary, the role of multiphoton ionization (MPI) and inverse Bremsstrahlung (IB) absorption of lasers in formation of plasma during LIBS has been discussed. A simple analysis of double pulse LIBS is also presented, which shows that delay between lasers play an important role in enhancement of LIBS intensity. The profile of enhancement in emission is mainly decided by electron-ion collision frequency in the plasma. The peak emission shifts to lower $\Delta t$ for higher value of $\upsilon_{ei}$, whereas profile gets broadened for lower value of $\upsilon_{ei}$. The correlation of experimental observations [11, 22] with theoretical results indicates that ablation after the second laser pulse increases $\geq 3.5$ times in comparison to single laser pulse ablation. It seems that drastic increase in mass ablated from the target material after the second laser pulse seems to be playing dominant role in deciding the magnitude of enhancement in emission. The change in the volume of emitting plasma under the effect of temperature or confinement effects (Magnetic or Spatial confinement of plasma) also affects the plasma emission. The plasma density and temperature affects the electron-ion collision frequency, which decides the threshold for any saturation in LIBS emission as $v_{ei} > 10^{13}$ s$^{-1}$. The effects of plasma shielding also contributes towards saturation in plasma emission. Finally results of these studies show that intensity of plasma emission can be increased and optimized for better sensitivity of LIBS only after carefully choosing the plasma parameters and the experimental conditions.

**FIGURE CAPTIONS**

1.  Variation in enhancement in emission with change in delay between lasers for different electron-ion collision frequency $\upsilon_{ei}$ ( $\dot{m}_1 = \dot{m}_2$ and $C_{s1} = C_{s2}$) [Ref. # 22].

2.  Variation in enhancement in emission with delay between lasers for different ratios of mass ablation rates $\dot{m}_2 / \dot{m}_1$ ($\upsilon_{ei} = 10^{10} \text{ s}^{-1}$, $C_{s1} = C_{s2}$) [Ref. # 22].

3.  Variation in enhancement in emission with time (detector gate delay) after second laser pulse for different electron–ion collision frequency $\upsilon_{ei}$ ( $\dot{m}_2 / \dot{m}_1 = 3.5$, $\Delta t = 3$ $\mu$s and $C_{s1} = C_{s2}$) [Ref. # 22].

4.  Variation in enhancement in emission with delay between lasers for different plasma expansion speed $C_{s2}/C_{s1}$ ($\upsilon_{ei} = 3.62 \times 10^{10} \text{ s}^{-1}$, $\dot{m}_1 = \dot{m}_2$) [Ref. # 22].

5.  Variation in enhancement in emission intensity with change in $C_{s2}/C_{s1}$ for lasers with different time duration [Ref. # 23].

6.  Variation in enhancement in emission intensity with an increase in plasma density for lasers of different time duration [Ref. # 23].

7.  Variation in enhancement in emission intensity with an increase in plasma temperature for lasers of different time duration [Ref. # 23].

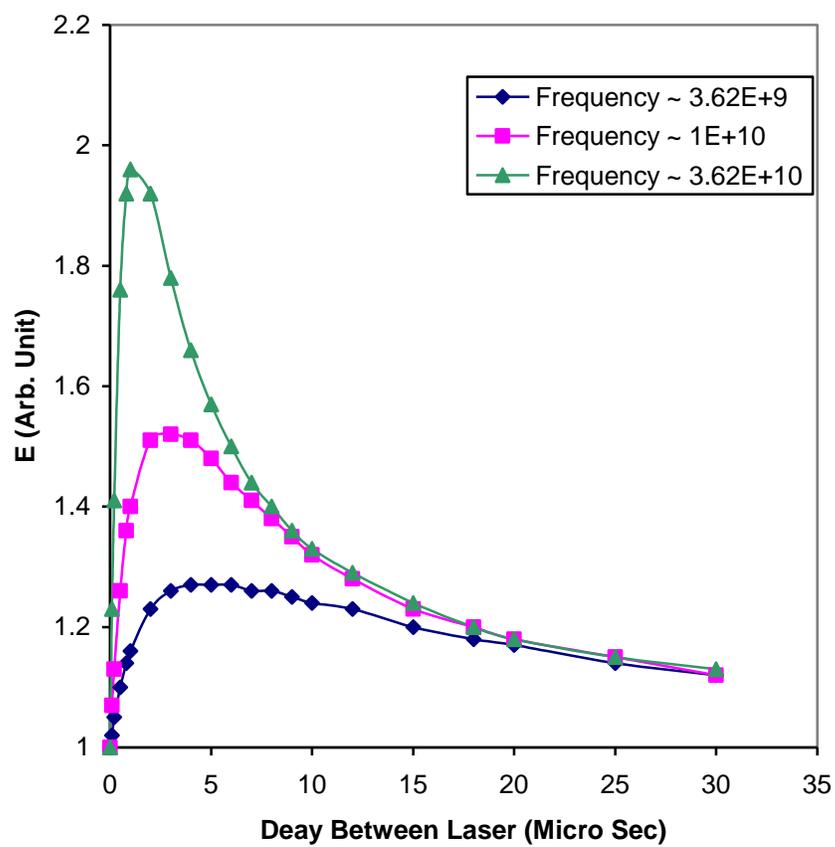

**Fig.-1**

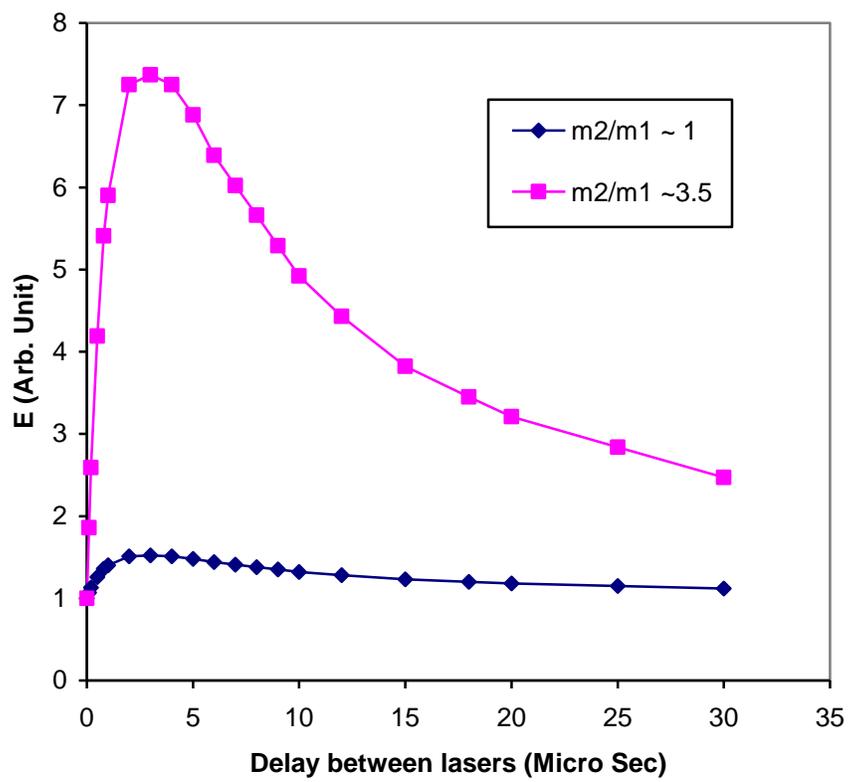

**Fig.-2**

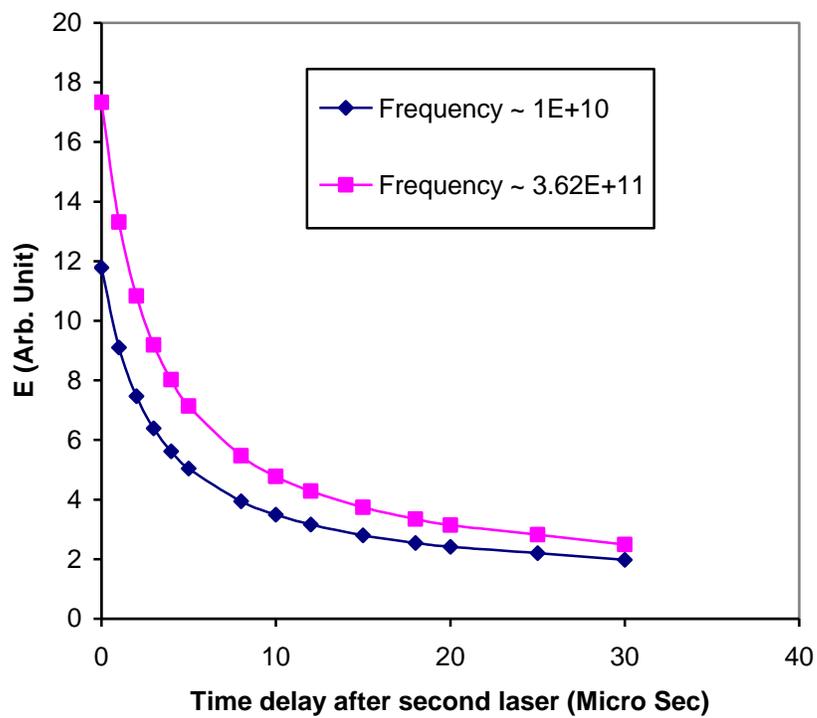

**Fig.-3**

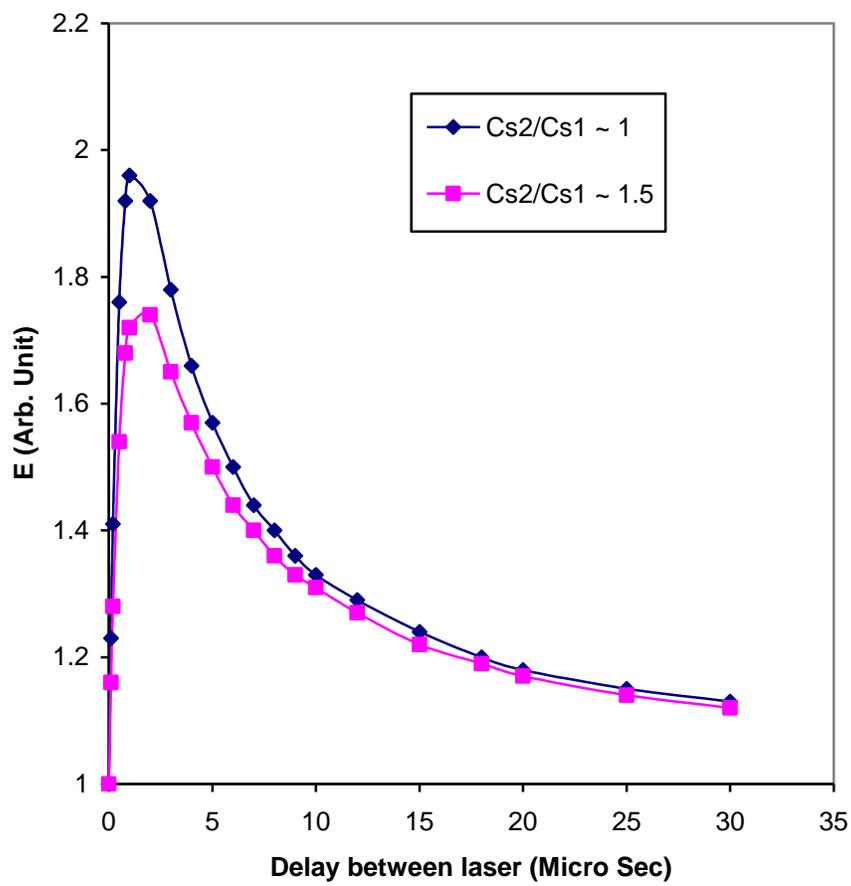

**Fig.-4**

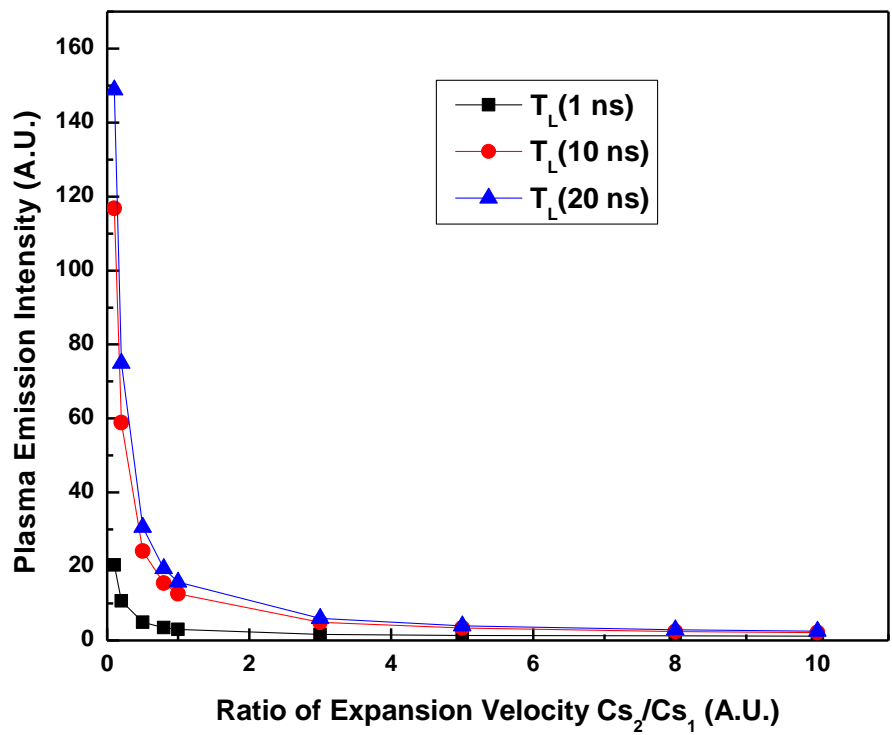

**Fig.- 5**

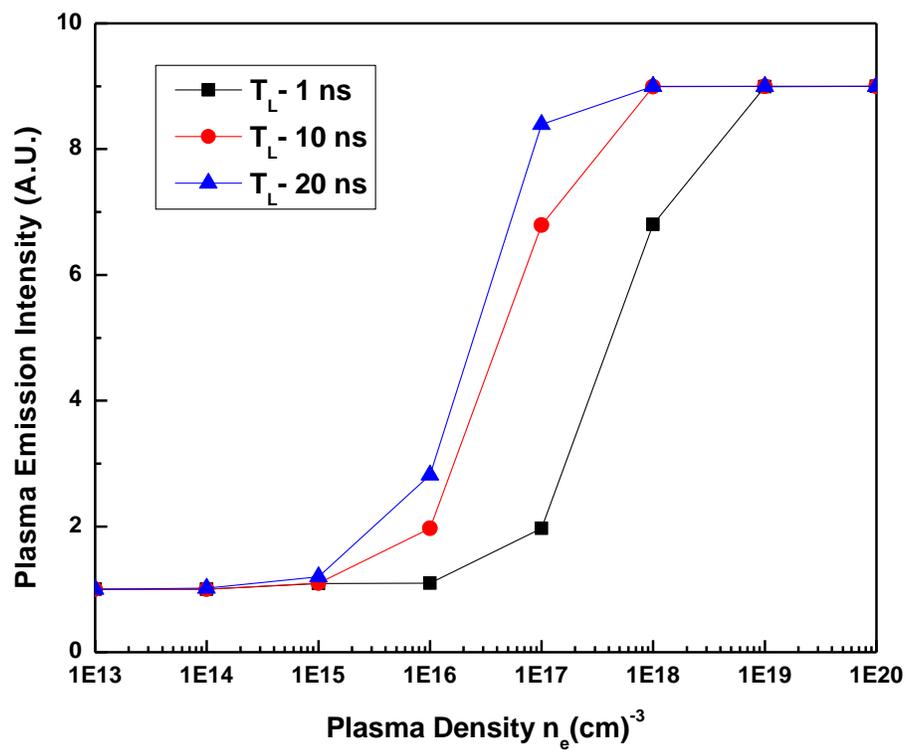

**Fig.-6**

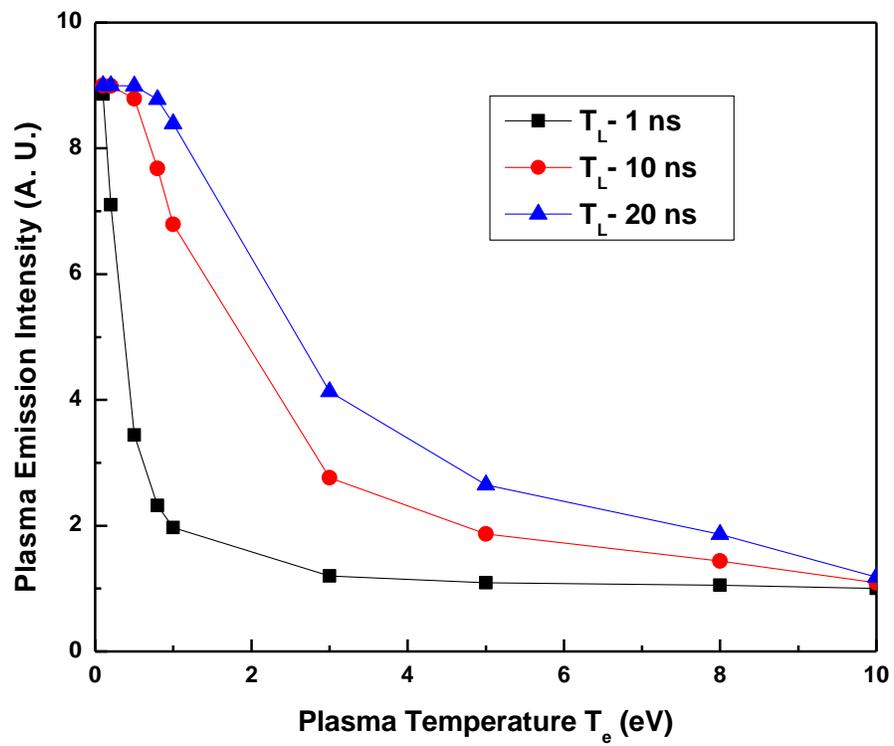

**Fig.-7**